\documentclass[twocolumn]{revtex4}
\usepackage[dvips]{graphicx}
\usepackage{amsfonts}
\usepackage{amsmath}
\usepackage{float, color,array, multirow}

\begin{document}

\title{Half-quantum flux in spin-triplet superconducting rings with bias current}

\author{Kazushi Aoyama}

\date{\today}

\affiliation{Department of Earth and Space Science, Graduate School of Science, Osaka University, Osaka 560-0043, Japan}

\begin{abstract}
Effects of a bias electric current have been theoretically investigated in a spin-triplet superconducting ring in a magnetic field. Based on the Ginzburg-Landau theory, we show that the bias current can stabilize a half-quantum-flux (HQF) state via couplings to the Zeeman field and the dipole-type spin-orbit interaction, the latter becoming active when the field is tilted from the ring axis. The emergence of the HQF state is reflected as a field-induced half-quantum-shift in the Little-Parks (LP) oscillation in the critical current. Possible relevance to recent LP experiments is also discussed. 
\end{abstract}

\maketitle

The macroscopic quantum nature of superconductivity is closely related to a phase of the superconducting (SC) gap function, one typical example of which is the Little-Parks (LP) oscillation appearing in a SC ring in a magnetic field ${\bf H}$ where a SC transition temperature $T_c$ exhibits a quantum oscillation as a function of field with its period being characterized by the flux quantum $\Phi_0=\frac{hc}{2|e|}$ \cite{LP_LittleParks_prl_62,LP_ParksLittle_pr_64,LP_Groff_pr_68}. When a bias electric current is further applied to the ring, the critical bias current $j_c$ and the associated resistivity also exhibit the LP oscillation \cite{LP-jc_Gurtovoi_jetp_07, LP-jc_Tokuda_jjap_22, LP_KA_prb_22}. In both the $T_c$ and $j_c$ cases, the oscillation peaks appear at $\Phi=n\Phi_0$, where $\Phi$ is the magnetic flux passing through the ring and $n$ is an integer corresponding to a winding number of the phase. Recently, half-quantum-shifted oscillations have been observed in current-biased LP experiments on the spin-triplet candidate $\beta$-Bi$_2$Pd \cite{Bi2Pd_Li_science_19} and the Bi/Ni bilayer as well \cite{Tokuda_private_23}, pointing to the emergence of half-quantum-flux (HQF) states. In the former $\beta$-Bi$_2$Pd polycrystalline samples, the $\Phi_0/2$ shift occurs at ${\bf H}=0$ due to the effect of grain boundaries, whereas in the latter homogeneous system, it occurs at a nonzero field ${\bf H}\neq 0$. In this letter, to gain insight into the origin of the field-induced switching into the HQF state, we theoretically investigate effects of the bias current on spin-triplet SC rings, putting an emphasis on the LP oscillation in $j_c$.      

\begin{figure}[t]
\begin{center}
\includegraphics[width=\columnwidth]{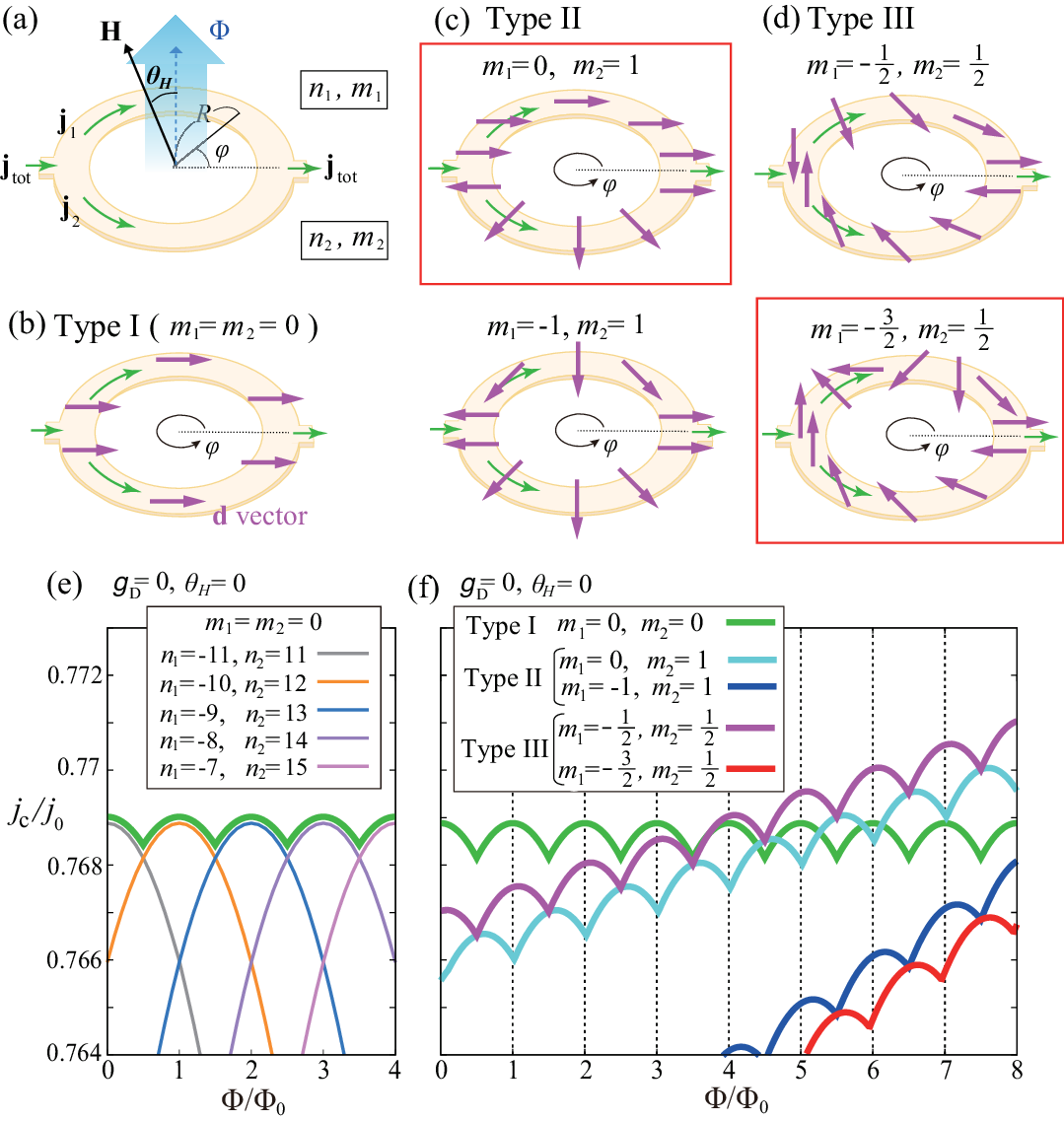}
\caption{(a) System setup; a spin-triplet SC ring in the presence of a bias electric current ${\bf j}_{\rm tot}={\bf j}_1+{\bf j}_2$ and a magnetic field ${\bf H}$ tilted by angle $\theta_H$ from the ring axis. 
(b)-(d) Schematically-drawn ${\bf d}$-vector textures \cite{LP_KA_prb_22}, where a magenta arrow represents the ${\bf d}$-vector. In (c) and (d), textured states enclosed by red rectangles correspond to the HQF states (for details, see the text). (e) and (f) Calculated critical bias current $j_c$ obtained at $T/T_{c0}=0.6$ for $R/\xi_0=50$ and $\eta=0.005$ in the {\it absence} of a spin-orbit coupling ($g_{\rm D}=0$) and the field tilt ($\theta_H=0$). (e) $j_c$ for the type-I state with $m_1=m_2=0$, where convex curves are obtained for given sets of $n_1$ and $n_2$, and their largest values are traced by a thick green curve which exhibits the Little-Parks (LP) oscillation. (f) $j_c$'s for the five types of ${\bf d}$-vector textures shown in (b)-(d), where the green curve is exactly the same as the one in (e) and the dashed lines indicate peak positions of the LP oscillation for the type-I state (the green curve). The HQF states are not realized for $g_{\rm D}=0$ and $\theta_H=0$. 
\label{fig:fig1}}
\end{center}
\end{figure}

In the conventional case of the spin-single $s$-wave ring, the LP oscillation in $T_c$ is usually explained by assuming that only the azimuthal angle $\varphi$ is relevant to the spatial variation of the SC gap function, i.e., $\Delta_s =|\Delta|e^{-i n \varphi}$ \cite{Thinkham}. Then, the SC current circulating around the ring is given by ${\bf j}\propto \hat{\varphi}|\Delta|^2(n-\frac{\Phi}{\Phi_0})$, where the first and second terms originate from the phase gradient and the London screening, respectively \cite{LP_ParksLittle_pr_64, Thinkham}. With increasing ${\bf H}$ or equivalently $\Phi$, $n$ switches to a larger integer to reduce ${\bf j}$. Such discontinuous change in $n$ emerges as the LP oscillation in $T_c$. In the presence of an additional bias current [see Fig. \ref{fig:fig1} (a)], upper and lower arms of the ring are not equivalent any more, each carrying a different phase-gradient current, so that they should possess different phase winding numbers $n_1$ and $n_2$. The LP oscillation in the critical bias current $j_c$ reflects discontinuous changes in the combination of $n_1$ and $n_2$ \cite{LP_KA_prb_22}. 

In the spin-triplet case of our interest, there exist not only phase but also spin degrees of freedom, the so-called ${\bf d}$-vector ${\bf d}=(d_x, d_y, d_z)$ which is related to the SC gap function via $\left( \Delta_{\uparrow\uparrow}, \Delta_{\uparrow\downarrow}, \Delta_{\downarrow\downarrow} \right) = \left( -d_x+id_y, d_z, d_x+id_y \right)$.
In a magnetic field as in the LP experiment, the $\uparrow\downarrow$ Cooper pair becomes unfavorable, so that the ${\bf d}$-vector tends to orient perpendicularly to the field, i.e., $d_z=0$ \cite{VW}. Then, the ${\bf d}$-vector on the ring without the bias current can be expressed as 
\begin{equation}\label{eq:HQV_general}
{\bf d} = \Big[ \frac{|\Delta_{\uparrow\uparrow}|}{\sqrt{2}} {\hat e}^+ \, e^{-i m \varphi} + \frac{|\Delta_{\downarrow\downarrow}|}{\sqrt{2}} {\hat e}^- \, e^{im \varphi}\Big] \, e^{-i n \varphi}
\end{equation}
with ${\hat e}^\pm =\frac{1}{\sqrt{2}}({\hat e}_x \pm i \, {\hat e}_y)$. Here, ${\hat e}_x$, ${\hat e}_y$, and $\hat{e}_z$ are mutually orthogonal unit-vectors in the spin space, and $\hat{e}_z$ corresponds to the quantization axis of the spin. Equation (\ref{eq:HQV_general}) is a SC analog of vortex states in superfluid $^3$He \cite{3He_vortex_Salomaa_rmp}, a prototype of the spin-triplet Cooper-pair condensate. To grasp the physical meanings of $m$ and $n$, it is convenient to consider the simplified case of $|\Delta|=|\Delta_{\uparrow\uparrow}|=|\Delta_{\downarrow\downarrow}|$ where Eq. (\ref{eq:HQV_general}) is reduced to ${\bf d} = |\Delta| \, e^{-i n \varphi} \, \big[ {\hat e}_x\cos(m \varphi) + {\hat e}_y\sin(m \varphi) \big]$. It is clear that $n$ is the phase winding number, whereas $m$ describes the rotation of the ${\bf d}$-vector along the circumference of the ring. For $m=0$, the ${\bf d}$-vector is spatially uniform, and $n$ takes integer values as in the spin-singlet case, whereas for $m=\pm 1/2$, the ${\bf d}$-vector rotates, and $n$ takes half-integer values since the gap function can also be single-valued for this choice of $m$ and $n$. The latter involves the HQF of $\Phi_0/2$ similarly to the half-quantum vortex in superfluid $^3$He \cite{HQV_ABM_Salomaa_prl_85, HQV_polar_Autti_prl_16, HQV_review_Sals_16, HQV_polar_Nagamura_prb_18}. This HQF state accompanied by the ${\bf d}$-vector rotation is usually unstable against the integer-flux state with $m=0$ and $n \in \mathbb{Z}$ due to the energy cost to form the ${\bf d}$-vector texture. 

In the presence of the bias current, the winding numbers in the upper arm $n_1$ and $m_1$ do not have to be the same as those in the lower arm $n_2$ and $m_2$, as in the spin-singlet case. Previously, we theoretically showed that various types of ${\bf d}$-vector textures with $m_1 \neq 0$ and/or $m_2 \neq 0$ can be stabilized by a combined effect of the Zeeman field and the bias current \cite{LP_KA_prb_22}. Even in this case, the HQF state is always unstable and the resultant LP oscillation in $j_c$ is not half-quantum-shifted [see Fig. \ref{fig:fig1} (f) and the text below]. 
In this letter, by introducing an additional dipole-type spin-orbit coupling which turns out to be active only when the external field is tilted from the ring axis [$\theta_H \neq 0$ in Fig. \ref{fig:fig1} (a)], we will demonstrate that a transition into the HQF state occurs at ${\bf H}\neq 0$, manifesting itself as a field-induced half-quantum-shift in the LP oscillation in $j_c$, as observed in the Bi/Ni-bilayer superconductor \cite{Tokuda_private_23}.

As a simple model, here, we consider the spin-triplet superconductor with the isotropic $p$-wave pairing in which the ${\bf d}$-vector is expressed as $d_\mu = \sum_i A_{\mu,i}\hat{p}_i$ with the order parameter $A_{\mu,i}$. The associated Ginzburg-Landau (GL) free energy functional ${\cal F}_{\rm GL}$ in a magnetic field is given by ${\cal F}_{\rm GL}= N(0)\int d{\bf r} ( f_0+f_{\rm D})$ with $f_0= f^{(2)}+\delta f^{(2)}+f^{(4)}$, where
\begin{eqnarray}\label{eq:GL}
f^{(2)} &=& \sum_{\mu,i,j} A^\ast_{\mu,i}\Big[ \frac{\alpha}{3}\delta_{i,j}A_{\mu,j} + K'\big( \Pi_j\Pi_j A_{\mu,i} + 2 \Pi_i \Pi_j A_{\mu,j} \big)\Big], \nonumber\\
\delta f^{(2)} &=& - \frac{\delta \alpha}{3} i \, \sum_{i} (A_{x,i}A^\ast_{y,i}-A_{y,i}A^\ast_{x,i}), \nonumber\\
f^{(4)} &=& \sum_{\mu,\nu,i,j} \big(\beta_1 |A_{\mu,i}A_{\mu,i}|^2+\beta_2(A^\ast_{\mu,i}A_{\mu,i})^2 \nonumber\\
&& + \beta_3 A^\ast_{\mu,i}A^\ast_{\nu,i}A_{\mu,j}A_{\nu,j} + \beta_4 A^\ast_{\mu,i}A_{\nu,i}A^\ast_{\nu,j}A_{\mu,j} \nonumber\\
&& + \beta_5 A^\ast_{\mu,i}A_{\nu,i}A_{\nu,j}A^\ast_{\mu,j} \big), \\
f_{\rm D} &=& g_{\rm D} \sum_{\mu,\nu} \big( A^\ast_{\mu,\mu} A_{\nu,\nu} + A^\ast_{\mu,\nu} A_{\nu,\mu} -\frac{2}{3}A^\ast_{\mu,\nu} A_{\mu,\nu} \big) . \nonumber
\end{eqnarray}
Equation (\ref{eq:GL}) is obtained from the GL free energy for the superfluid $^3$He by the replacement $-i {\boldsymbol\nabla} \rightarrow {\boldsymbol\Pi}=-i {\boldsymbol\nabla}+2|e|{\bf A}$ with the vector potential ${\bf A}=\frac{1}{2}{\bf H}\times{\bf r}$, so that in the weak-coupling limit, the coefficients are given by $\alpha=\ln\frac{T}{T_{c0}}$, $K'=\frac{1}{5}(\frac{T_{c0}}{T})^2 \, \xi_0^2$, $-2\beta_1 = \beta_2=\beta_3=\beta_4=-\beta_5 = 2\beta_0= \frac{7\zeta(3)}{120 \pi^2T^2}$ with the SC transition temperature at zero field $T_{c0}$ and the SC coherence length $\xi_0=\sqrt{\frac{7\zeta(3)}{48 \pi^2}} \, \frac{v_{\rm F}}{T_{c0}}$ \cite{VW}. The $\delta f^{(2)}$ term originates from the difference in density of states between the up-spin and down-spin Fermi surfaces \cite{VW,HQV_Vakaryuk_prl_09}, so that within a linear approximation, we have $\delta \alpha \propto |{\bf H}|$. Density of states averaged over the two Zeeman-split Fermi surfaces is denoted by $N(0)$. The additional term which is not incorporated in the previous work \cite{LP_KA_prb_22} is the dipole interaction $f_{\rm D}$. It represents a kind of spin-orbit coupling in the sense that the spin and orbital subscripts $\mu$ and $i$ in $A_{\mu,i}$ are coupled in $f_{\rm D}$. This dipole-type spin-orbit coupling is known to be important for texture formations in superfluid $^3$He \cite{VW}, and such a situation is also the case for the present system, as will be discussed below. 

Figure \ref{fig:fig1} (a) shows the system setup, a SC ring of mean radius $R$ in a magnetic field tilted from the ring axis by angle $\theta_H$. When the ring-width $w$ and thickness $d$ are sufficiently small, being comparable to $\xi_0$, the vector potential can be expressed as ${\bf A}=\frac{HR\cos\theta_H}{2}\hat{\varphi}$, so that we have ${\boldsymbol\Pi}= \hat{\varphi}\frac{1}{R} \big( -i \partial_\varphi + \frac{\Phi}{\Phi_0}\big)$ with $\Phi=\pi R^2 H \cos\theta_H$. Also, the effect of the Zeeman splitting $\delta \alpha$ can be expressed as $\delta \alpha = \eta \, (\Phi/\cos\theta_H)/\Phi_0$. Note that for $\theta_H \neq 0$, the spin quantization axis $\hat{e}_z$ is tilted from the ring axis. 

Among various possible symmetries of $A_{\mu, i}$, we consider axially-symmetric states compatible with the ring geometry. A typical example of such axial states is the Anderson-Brinkman-Morel state of the form $A_{\mu,i}=d_\mu \, w_{{\bf p},i}$ with $w_{{\bf p},i}=\frac{1}{\sqrt{2}}(\hat{p}_x+i \, \hat{p}_y)_i$, which is known to be realized in the superfluid $^3$He A phase \cite{VW}. In the present system, we assume that $\hat{p}_z$ is parallel to the ring axis. Since the dipole energy is calculated as $f_{\rm D} =g_{\rm D}(\frac{1}{3}|{\bf d}|^2-|{\bf d}\cdot \hat{p}_z|^2)$, it tends to orient the ${\bf d}$-vector parallel to the $\hat{p}_z$ direction (ring axis), which is a manifestation of the spin-orbit coupling nature of $f_{\rm D}$. We note that the following results are qualitatively unchanged even if the polar pairing state of $w_{{\bf p},i}=\hat{p}_z$ is assumed \cite{KA_supple}, and thus, the chiral nature of the orbital state is not important.

In the presence of the bias current, 
Eq. (\ref{eq:HQV_general}) can be extended to \cite{LP_KA_prb_22}
\begin{equation}\label{eq:OP_t}
d_\mu = \left\{\begin{array}{l}
\big[\frac{|\Delta_{\uparrow\uparrow}|}{\sqrt{2}} {\hat e}^+_\mu  e^{-i m_1 \varphi}+ \frac{|\Delta_{\downarrow\downarrow}|}{\sqrt{2}} {\hat e}^-_\mu  e^{i m_1 \varphi}\big] e^{-i n_1 \varphi}\, (0\leq \varphi \leq \pi) \nonumber\\
\big[\frac{|\Delta_{\uparrow\uparrow}|}{\sqrt{2}} {\hat e}^+_\mu  e^{-i m_2 \varphi} + \frac{|\Delta_{\downarrow\downarrow}|}{\sqrt{2}}{\hat e}^-_\mu  e^{i m_2 \varphi}\big]  e^{-i n_2 \varphi} \, (\pi \leq \varphi \leq 2\pi) 
\end{array} \right. .
\end{equation}
The upper-arm winding numbers $n_1$ and $m_1$ are correlated to the lower-arm ones $n_2$ and $m_2$ via a constraint to avoid sign mismatches at the intersections between the two arms, namely, at $\varphi=0$ and $\pi$; $n_1-n_2$ must be an even (odd) integer when $m_1-m_2$ is an even (odd) integer. 

Following Ref. \cite{LP_KA_prb_22}, the usual integer-flux state with the uniform ${\bf d}$-vector ($m_1=m_2=0$) will be called type I, whereas  the states possessing nonzero integer and half-integer values of $m_1$ and $m_2$ will be called type II and III, respectively. These three types of the flux states are shown in Figs. \ref{fig:fig1} (b)-(d) where $|\Delta_{\uparrow\uparrow}|=|\Delta_{\downarrow\downarrow}|=|\Delta|$ is assumed for ease of understanding and then, the SC current ${\bf j}=-\frac{\delta {\cal F}_{\rm GL}}{\delta {\bf A}}=- 8N(0)|e|K'   \sum_{\mu} d^\ast_\mu {\boldsymbol\Pi} d_\mu$ satisfies the relation $\oint {\bf j}\cdot d{\boldsymbol\varphi} \propto 2\pi |\Delta|^2\big( \frac{n_1+n_2}{2}-\frac{\Phi}{\Phi_0} \big)$(for the concrete expression of ${\bf j}$ in each arm, see below). Thus, among the type-II and -III states, the specific combinations of $n_1$, $n_2$, $m_1$, and $m_2$ having half-integer values of $(n_1+n_2)/2$  correspond to the HQF state having a half-integer $\Phi/\Phi_0$. For half-integer $(n_1+n_2)/2$, $(m_1+m_2)/2$ must also be a half integer because of the constraint on $n_1$, $n_2$, $m_1$, and $m_2$, and thus, it turns out that the textured states enclosed by red rectangles in Figs. \ref{fig:fig1} (c) and (d) correspond to the HQF state. 

Now, we will discuss the critical bias current $j_c$ \cite{LP_KA_prb_22}. With the use of the above expression for $d_\mu$, the SC current in each arm ${\bf j}_l$ ($l$=1,2) can be calculated as ${\bf j}_{\, l}=(-1)^l (4N(0)|e|K'/R) \big[ |\Delta_{\uparrow\uparrow}|^2(n_l+m_l-\frac{\Phi}{\Phi_0})+\ |\Delta_{\downarrow\downarrow}|^2(n_l-m_l-\frac{\Phi}{\Phi_0})\big]\hat{\varphi}$, so that the total bias current ${\bf j}_{\rm tot}={\bf j}_1+{\bf j}_2$ is given by
\begin{eqnarray}\label{eq:current_tot_t}
{\bf j}_{\rm tot} &=&-{\hat \varphi} \, \frac{4N(0)|e|K'}{R}\Big[(|\Delta_{\uparrow\uparrow}|^2+|\Delta_{\downarrow\downarrow}|^2)(n_1-n_2) \nonumber\\
&+& (|\Delta_{\uparrow\uparrow}|^2-|\Delta_{\downarrow\downarrow}|^2)(m_1-m_2) \Big] .
\end{eqnarray}
One can see from Eq. (\ref{eq:current_tot_t}) that the difference in the phase winding number $n_{\rm tot}=n_2-n_1$ corresponds to the bias current ${\bf j}_{\rm tot}$, and the ${\bf d}$-vector texture yields an additional contribution proportional to $m_1-m_2$. The free energy ${\cal F}_{\rm GL}=N(0) Rwd \, \big(\int_0^\pi + \int_\pi^{2\pi} \big) d\varphi \,[ f_0+f_{\rm D}]$ is also a function of the four winding numbers, so that by substituting $|\Delta_{\sigma\sigma}|$ determined by the GL equation $\frac{\delta {\cal F}_{\rm GL}}{\delta |\Delta_{\sigma\sigma}|}=0$ into Eq. (\ref{eq:current_tot_t}), we can calculate $|{\bf j}_{\rm tot}|$'s for various combinations of $n_1$, $n_2$, $m_1$, and $m_2$ among which the largest one gives $j_c$. 

Figures \ref{fig:fig1} (e) and (f) show the results for $g_{\rm D}=\theta_H=0$, where other parameters are set to be $T/T_{c0}=0.6$, $R/\xi_0=50$, and $\eta=0.005$, and $j_c$ is normalized by $j_0=N(0)T_{c0}|e|v_{\rm F}$. This case of $g_{\rm D}=\theta_H=0$, which has already been discussed in Ref. \cite{LP_KA_prb_22}, is picked up here just for reference. In the case of the type-I state shown in Fig. \ref{fig:fig1} (e), among $j_c$'s obtained for different combinations of $n_1$ and $n_2$, the highest ones indicated by the green curve exhibit the conventional LP oscillation with its peaks at integer $\Phi/\Phi_0$. The textured states of $m_1 - m_2 \neq 0$, on the other hand, can yield a higher $j_c$ which, due to the second term in Eq. (\ref{eq:current_tot_t}), increases almost linearly in $H$ as $|\Delta_{\uparrow\uparrow}|^2-|\Delta_{\downarrow\downarrow}|^2$ can roughly be estimated as $\sim \eta H$ \cite{LP_KA_prb_22}. Actually, as shown in Fig. \ref{fig:fig1} (f), the type-III state with $m_1=-m_2=-1/2$ is realized at high fields, giving the highest $j_c$ (see the violet curve). Even in this type-III state,  the phase of the LP oscillation remains unchanged. Although the half-quantum-shifted LP oscillations are potentially possible (see the cyan and red curves), such HQF states are not realized as their $j_c$'s are always smaller than those of the counter integer-flux states with $m_1+m_2=0$ (see the violet and blue curves). 

It is useful to understand, from the view point of the free energy, the reason why the HQF states are always unstable. In the case of $g_{\rm D}=\theta_H=0$, the free energy density averaged over the whole ring $\overline{f_0}=\frac{1}{2\pi}\big(\int_0^\pi+\int_\pi^{2\pi} \big) d\varphi \,f_0$ reads \cite{LP_KA_prb_22}
\begin{eqnarray}\label{eq:f0}
&&\overline{f_0}= \frac{\alpha-\delta \alpha}{6}|\Delta_{\uparrow\uparrow}|^2 + \frac{\alpha+\delta \alpha}{6}|\Delta_{\downarrow\downarrow}|^2 +  \beta_0 \big(|\Delta_{\uparrow\uparrow}|^4+|\Delta_{\downarrow\downarrow}|^4\big) \nonumber\\
&&+ \frac{K'}{R^2} \big(|\Delta_{\uparrow\uparrow}|^2+|\Delta_{\downarrow\downarrow}|^2 \big) \Big[\Big(n_2-\frac{n_{\rm tot}}{2}-\frac{\Phi}{\Phi_0}+\lambda\frac{m_1+m_2}{2}\Big)^2 \nonumber\\
&& +\frac{n_{\rm tot}^2}{4}+\frac{m_1^2}{2}+\frac{m_2^2}{2} -  \lambda \, n_{\rm tot} \frac{m_1-m_2}{2} -\Big(\lambda \frac{m_1+m_2}{2}\Big)^2\Big] 
\end{eqnarray}
with $\lambda = \frac{|\Delta_{\uparrow\uparrow}|^2-|\Delta_{\downarrow\downarrow}|^2}{|\Delta_{\uparrow\uparrow}|^2+|\Delta_{\downarrow\downarrow}|^2}$.
Due to the $\lambda  \,n_{\rm tot} \, (m_1-m_2)$ term which is roughly proportional to $\eta H \, |{\bf j}_{\rm tot}| (m_1-m_2)$, the ${\bf d}$-vector texture with a larger $m_1-m_2 \neq 0$ can acquire a larger energy gain. This situation is analogous to noncentrosymmetric superconductors in a magnetic field \cite{NCS_book} where a similar field-current coupling term induces the phase-modulated helical SC state \cite{SC_jM,SC_Bj,Dimitrova,Samokhin,Kaur,Fujimoto,Helical_AS_12,Helical_ASS_14}.   
In Eq. (\ref{eq:f0}), for a fixed value of $m_1-m_2$, the energy cost for the texture formation $m_1^2+m_2^2$ takes a minimum value at $m_1+m_2=0$, so that the HQF state with a half-integer $(m_1+m_2)/2\neq 0$ is unfavorable compared with the counter integer-flux state with $m_1+m_2=0$. 

\begin{figure*}[t]
\begin{center}
\includegraphics[scale=0.48]{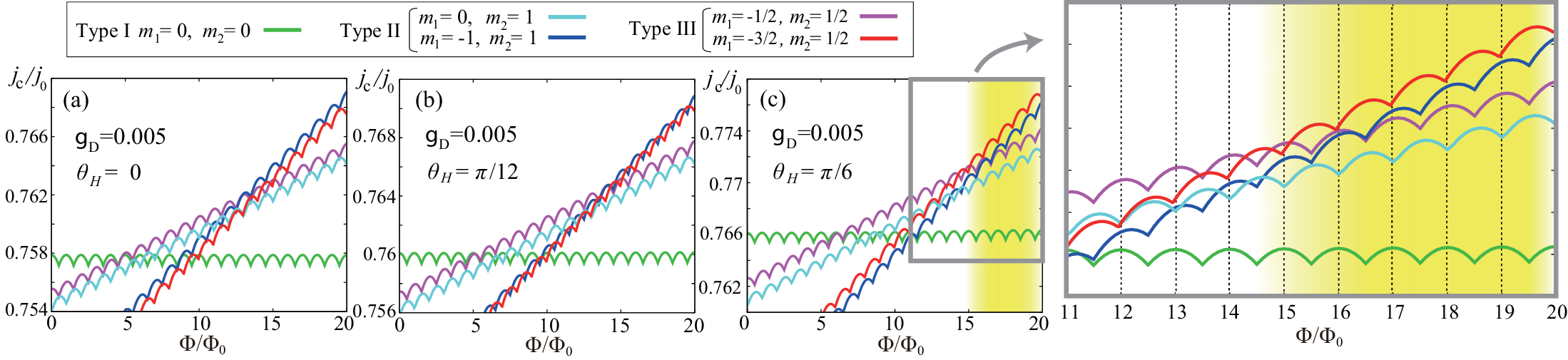}
\caption{The LP oscillation in $j_c$ obtained for $g_{\rm D}=0.005$. (a)-(c) The results for $\theta_H=0$, $\pi/12$, and $\pi/6$, respectively, where other parameters ($T/T_{c0}=0.6$, $R/\xi_0=50$, and $\eta=0.005$) and color notations are the same as those in Fig. \ref{fig:fig1} (f). In (c), a right panel shows a zoomed view of the region enclosed by a rectangle in the left panel. 
The HQF state with $m_1=-\frac{3}{2}$ and $m_2=\frac{1}{2}$ [see Fig. \ref{fig:fig1} (d)] is realized at high fields (see the red curve in the yellow region), showing the half-quantum-shifted LP oscillation. \label{fig:fig2}}
\end{center}
\end{figure*}

The above situation, however, can be changed by the dipole-type spin-orbit interaction whose energy density $\overline{f_{\rm D}}=\frac{1}{2\pi}\big(\int_0^\pi+\int_\pi^{2\pi} \big) d\varphi \,f_{\rm D}$ can be evaluated \cite{KA_supple} as  
\begin{equation}\label{eq:energy_dipole}
\overline{f_{\rm D}} = -g_{\rm D}\Big[ \big(|\Delta_{\uparrow\uparrow}|^2+|\Delta_{\downarrow\downarrow}|^2\big)c+|\Delta_{\uparrow\uparrow}||\Delta_{\downarrow\downarrow}|\frac{c_{1}+c_{2}}{2}\Big] \nonumber
\end{equation}
with $c = (3\sin^2\theta_H-2)/12$ and
\begin{equation}
\frac{c_1+c_2}{2}=\frac{\sin^2\theta_H}{4} \Big\{ (\delta_{m_1,0}+\delta_{m_2,0}) + \Big|\frac{\delta_{2m_1, {\rm odd}}}{\pi \, m_1 }-\frac{\delta_{2m_2, {\rm odd}}}{\pi \, m_2 } \Big|\Big\}. \nonumber
\end{equation}
The $|\Delta_{\uparrow\uparrow}||\Delta_{\downarrow\downarrow}|(c_1+c_2)$ term can lower the energy of the HQF state with $m_1=-\frac{3}{2}$ and $m_2=\frac{1}{2}$, keeping the energy of the counter integer-flux state with $m_1=-1$ and $m_2=1$ unchanged, so that it can stabilize the HQF state, compensating the relative energy cost for the texture formation. This energy gain mechanism is active only when the field is tilted from the ring axis ($\theta_H\neq 0$) and the applied bias current breaks the symmetry of the upper and lower arms ($m_1 \neq m_2$).

Figures \ref{fig:fig2} (a)-(c) show the $\theta_H$ dependence of $j_c$ obtained for $g_{\rm D}=0.005$, where other parameters are the same as those for Fig. \ref{fig:fig1} (f). With increasing $\theta_H$, $j_c$ for the type-III HQF state with $m_1=-\frac{3}{2}$ and $m_2=\frac{1}{2}$ (red curve) is gradually elevated to overwhelm $j_c$ for the type-II integer-flux state with $m_1=-1$ and $m_2=1$ (blue curve). As a result, in the case of $\theta_H=\pi/6$ shown in Fig. \ref{fig:fig2} (c), the HQF state gives the highest $j_c$ in the high-field region of $\Phi/\Phi_0>15$ (yellow region), exhibiting the half-quantum-shifted LP oscillation, which can clearly be seen in the zoomed view of Fig. \ref{fig:fig2} (c). Even for smaller $\theta_H$, a stronger spin-orbit coupling (larger $g_{\rm D}$) can stabilize the HQF state, as exemplified by Fig. 1 in Supplemental Material \cite{KA_supple}.  

In this letter, we have theoretically shown that in the spin-triplet SC ring, the applied bias current can stabilize the HQF state via the couplings to the Zeeman field and the dipole-type spin-orbit interaction, the latter effect becoming active when the field is tilted from the ring axis. It is also demonstrated that the emergence of the HQF state is reflected as the field-induced half-quantum-shift in the LP oscillation. In the recent LP experiment on the Bi/Ni bilayer superconductor whose pairing symmetry is still under discussion \cite{BiNi_SC_Gong_cpl_15, BiNi_SC_Gong_sadv_17, BiNi_SC_Ghadimi_prb_19, BiNi_SC_Chauhan_prl_19, BiNi_SC_Chao_prb_19}, a similar field-induced $\Phi_0/2$ shift has been observed \cite{Tokuda_private_23}, being distinguished from the zero-field $\Phi_0/2$ shift originating from the grain-boundary effect \cite{Bi2Pd_Li_science_19, Bi2Pd_Xu_prl_20}. 
In this experiment, an out-of-plane applied magnetic field and a weak in-plane stray field stemming from the Ni-layer spontaneous magnetization \cite{BiNi_mag_Zhou_jmmm_17} may serve as an effective tilted field of $\theta_H \neq 0$. 
Although in this case, $\theta_H$ should be dependent on the applied-field, in contrast to the situation considered here, i.e., $\theta_H={\rm const.}$, our result could be applied to the higher-field region of the Bi/Ni system where the effect of the applied out-of-plane field becomes more dominant than the effects of the in-plane magnetization and the dipole-type spin-orbit coupling, keeping $\theta_H$ to an almost-constant small value. The qualitative agreement between the theoretical and experimental results suggests that the spin-triplet HQF state may be realized in the Bi/Ni bilayer superconductor. It should be emphasized that in contrast to the well-known effect of the Fermi-liquid correction which suppresses the ratio of the spin superfluid and superfluid densities, resulting in the occurrence of the HQF (or half-quantum-vortex) states \cite{HQV_ABM_Salomaa_prl_85, HQV_polar_Nagamura_prb_18, HQV_Vakaryuk_prl_09} accompanied with a peak-split LP oscillation over the whole field range \cite{Sr2RuO4_HQF_Vakaryuk_prl_11,Sr2RuO4-LP_Cai_arXiv_20}, the mechanism presented here, i.e., the combined effect of the bias current and the dipole-type spin-orbit coupling, yields not such a peak-splitting but a field-induced switching to the $\pi$-shifted LP oscillation.

Of course, a dominant spin-orbit coupling and the resultant ${\bf d}$-vector locking generally depend on the details of specific systems. Actually, for the spin-triplet candidate Sr$_2$RuO$_4$ \cite{Sr2RuO4_review_Maeno_rmp_03} (its pairing symmetry is still controversial \cite{Sr2RuO4_NMR_Pustogow_nature_19}), an in-plane-field-induced ${\bf d}$-vector locking different from the dipole one is phenomenologically proposed \cite{Sr2RuO4_torque_Jang, Sr2RuO4_HQF_Vakaryuk_prl_11, Sr2RuO4-LP_Yasui_prb_17, Sr2RuO4-LP_Cai_arXiv_20}. Nevertheless, at least it is certain that for the micrometer- or submicrometer-sized small rings of our interest, the dipole-type spin-orbit coupling could be relevant. The field strength necessary to overcome the dipole-locking is about 30 G in the well-established case of superfluid $^3$He \cite{VW}. Since this characteristic field strength depends basically on the Fermi-surface structure but not on the size of the magnetic moment \cite{Hasegawa_dipole_jpsj_03}, a dipole field of a similar order is also expected in electron systems. In the case of Sr$_2$RuO$_4$, an estimation shows that it is about 200 G \cite{Miyake_dipole_jpsj_10}. For a small SC ring with diameter of, for example, 1 $\mu$m, the flux quantum $\Phi_0\sim 2.0 \times 10^{-15}$ Wb corresponds to about 20 G \cite{Bi2Pd_Li_science_19}, so that in the case of Fig. \ref{fig:fig2} (d), the transition into the HQF state should occur at $\Phi=15\Phi_0\sim 300$ G which is slightly larger than or comparable to the typical dipole-field, suggesting that the dipole-type spin-orbit coupling could be non-negligible for the LP physics in spin-triplet superconductors. Although a material-specific microscopic theory would be necessary to fully understand the SC properties, we believe that our result will lead to further exploration and understanding of spin-triplet superconductors. 

\begin{acknowledgments}
The author thanks M. Tokuda, Y. Niimi and T. Mizushima for useful discussions. This work is partially supported by JSPS KAKENHI Grant No. JP21K03469 and JP23H00257.
\end{acknowledgments}

\hspace{2cm}
\pagebreak

\onecolumngrid
\hspace{2cm}
\begin{center}
\textbf{\large Supplemental Material for ''Half-quantum flux in spin-triplet superconducting rings with bias current''}\\[.2cm]
\end{center}

In the main text, we show that the dipole-type spin-orbit coupling, which is denoted by $f_{\rm D}$ in the Ginzburg-Landau (GL) free energy functional [see Eq. (2) in the main text], favors a half-quantum-flux (HQF) state in the presence of a bias current and a magnetic field tilted from the ring axis by angle $\theta_H \neq 0$, where the chiral orbital state of the form $w_{{\bf p},i}=\frac{1}{\sqrt{2}}(\hat{p}_x+i\hat{p}_y)_i$ is assumed for the spin-triplet $p$-wave superconducting (SC) order parameter $A_{\mu,i}=d_\mu w_{{\bf p},i}$. In this supplemental material, starting from the derivation of the concrete expression of $f_{\rm D}$, we show that even if the polar state of $w_{{\bf p},i}=\hat{p}_{z,i}$ is assumed as another axial pairing state, the result discussed in the main text is qualitatively unchanged. 

\section{Concrete expression of the dipole-type spin-orbit coupling term}
We first derive the concrete expressions of $f_{\rm D}$ for the two pairing states, i.e., the chiral and polar states. 
In both pairing states, the free energy density ${\cal F}_{\rm GL}/[N(0)Rwd]=\overline{f_0}+\overline{f_{\rm D}}$ takes the same form as follows:

\begin{eqnarray}\label{eq:energy}
\overline{f_0} &=& \frac{1}{2}\sum_{\sigma=\uparrow, \downarrow}\bigg[ \frac{\alpha-\epsilon_{\sigma} \, \delta \alpha}{3}|\Delta_{\sigma\sigma}|^2  +  C_4 \beta_0 |\Delta_{\sigma\sigma}|^4 + C_K \frac{K'}{R^2}|\Delta_{\sigma\sigma}|^2\Big\{ \big(n_1+\epsilon_\sigma m_1-\frac{\Phi}{\Phi_0}\big)^2 + \big(n_2+\epsilon_\sigma m_2-\frac{\Phi}{\Phi_0}\big)^2\Big\} \bigg], \nonumber\\
\overline{f_{\rm D}} &=& -g_{\rm D}C_{\rm D} \Big[ \big(|\Delta_{\uparrow\uparrow}|^2+|\Delta_{\downarrow\downarrow}|^2\big)c+|\Delta_{\uparrow\uparrow}||\Delta_{\downarrow\downarrow}|\frac{c_{1}+c_{2}}{2}\Big], \nonumber\\ 
c &=& \frac{(\hat{e}_x\cdot\hat{p}_z)^2+(\hat{e}_y\cdot\hat{p}_z)^2}{4}-\frac{1}{6}, \nonumber\\
c_{l} &=& \frac{(\hat{e}_x\cdot\hat{p}_z)^2-(\hat{e}_y\cdot\hat{p}_z)^2}{2} {\rm Re}(\gamma_l) + (\hat{e}_x\cdot\hat{p}_z)(\hat{e}_y\cdot\hat{p}_z) {\rm Im}(\gamma_l) , \nonumber\\
\gamma_{l} &=& \frac{1}{\pi}\int_{(l-1)\pi}^{l\pi}d\varphi \, e^{2 m_l  i \varphi} =\delta_{m_l, 0} + i \frac{(-1)^{(l+1)}}{\pi m_l} \delta_{2m_l, {\rm odd}} ,
\end{eqnarray}
where $\epsilon_{\uparrow (\downarrow)}=1 \, (-1)$ and other notations are the same as those in the main text. 
The difference between the two pairing states consists only in the coefficients $C_4$, $C_K$, and $C_{\rm D}$ which are calculated as
\begin{equation}\label{eq:coefficient}
C_4=\left\{\begin{array}{c} 
2 \\
3 \end{array} ,\right. \quad C_K= \left\{\begin{array}{c} 
1 \\
1/2 \end{array} , \right. \quad C_{\rm D}= \left\{\begin{array}{c} 
1 \\
-2 \end{array}  \right.  \quad \left. \begin{array}{l} 
({\rm chiral:} \, w_{{\bf p},i} = \frac{1}{\sqrt{2}}(\hat{p}_x+i\hat{p}_y)_i )\\
({\rm polar:} \, w_{{\bf p},i} = \hat{p}_{z,i}) \end{array} \right. .
\end{equation}
Note that the difference in $C_K$ is directly related to the SC current as
\begin{equation}\label{eq:current}
{\bf j}_{\rm tot} = -{\hat \varphi} \, C_K  \frac{4|e|K'}{R}\Big[(|\Delta_{\uparrow\uparrow}|^2+|\Delta_{\downarrow\downarrow}|^2)(n_1-n_2) + (|\Delta_{\uparrow\uparrow}|^2-|\Delta_{\downarrow\downarrow}|^2)(m_1-m_2) \Big] .
\end{equation}

Although in Eq. (\ref{eq:coefficient}), the sign of the coefficient of the dipole energy $C_{\rm D}$ depends on the pairing symmetry,  this does not mean that in the chiral and polar states, the dipole interaction acts in opposite manners. Since $\overline{f_{\rm D}}$ depends on the relative angle between the ${\bf d}$-vector and the $\hat{p}_z$ direction, we consider the ${\bf d}$-vector configuration to minimize $\overline{f_{\rm D}}$. Noting that the spin-space triad $\hat{e}_x$, $\hat{e}_y$, and $\hat{e}_z$ can generally be expressed by the momentum-space (laboratory-frame) coordinates $\hat{p}_x$, $\hat{p}_y$, and $\hat{p}_z$ with the use of Euler angles, we have
\begin{eqnarray}
(\hat{e}_x\cdot\hat{p}_z)^2+(\hat{e}_y\cdot\hat{p}_z)^2 &=& \sin^2 \theta_H , \nonumber\\
(\hat{e}_x\cdot\hat{p}_z)^2-(\hat{e}_y\cdot\hat{p}_z)^2 &=& \sin^2 \theta_H \cos (2\psi), \nonumber\\
(\hat{e}_x\cdot\hat{p}_z)(\hat{e}_y\cdot\hat{p}_z) &=& -\frac{1}{2}\sin^2 \theta_H \sin (2\psi), 
\end{eqnarray}    
where $\theta_H$ is the tilting angle of the magnetic field and $\psi$ describes the ${\bf d}$-vector rotation within the $\hat{e}_x$-$\hat{e}_y$ plane perpendicular to the tilted field. Since the spin-space is assumed to be isotropic and thus, the ${\bf d}$-vector can freely rotate within the the $\hat{e}_x$-$\hat{e}_y$ plane, the rotation angle $\psi$ can be chosen such that the overall coefficient of the $|\Delta_{\uparrow\uparrow}||\Delta_{\downarrow\downarrow}|$ term in $\overline{f_{\rm D}}$ takes the smallest negative value. Noting that $(\hat{e}_x\cdot\hat{p}_z)^2-(\hat{e}_y\cdot\hat{p}_z)^2$ and $(\hat{e}_x\cdot\hat{p}_z)(\hat{e}_y\cdot\hat{p}_z)$ appear in the different sectors of $m_l$ [see $c_l$ and $\gamma_l$ in Eq. (\ref{eq:energy})], the minimization condition for $\overline{f_{\rm D}}$ turns out to be $\cos(2\psi)=-1$ (+1) and $\sin(2\psi)=1 \times {\rm sgn}[\frac{1}{m_1}-\frac{1}{m_2}]$  ($-1 \times {\rm sgn}[\frac{1}{m_1}-\frac{1}{m_2}]$) in the chiral (polar) pairing state. Thus, the HQF state with $m_1=-\frac{3}{2}$ and $m_1=\frac{1}{2}$ acquires the additional dipole-energy gain 
\begin{equation}\label{eq:energy-gain}
-g_{\rm D} C_{\rm D} |\Delta_{\uparrow\uparrow}||\Delta_{\downarrow\downarrow}| \frac{c_1+c_2}{2}= -g |C_{\rm D}|  |\Delta_{\uparrow\uparrow}||\Delta_{\downarrow\downarrow}|\frac{\sin^2\theta_H}{4} \Big|\frac{\delta_{2m_1, {\rm odd}}}{\pi m_1}-\frac{\delta_{2m_2, {\rm odd}}}{\pi m_2} \Big|  \nonumber
\end{equation}  
compared with the energetically competing integer-flux state with $m_1=-1$ and $m_2=1$, suggesting that the dipole-type spin-orbit coupling favors the HQF state in both the chiral and polar pairing states.

\begin{figure*}[t]
\includegraphics[scale=0.48]{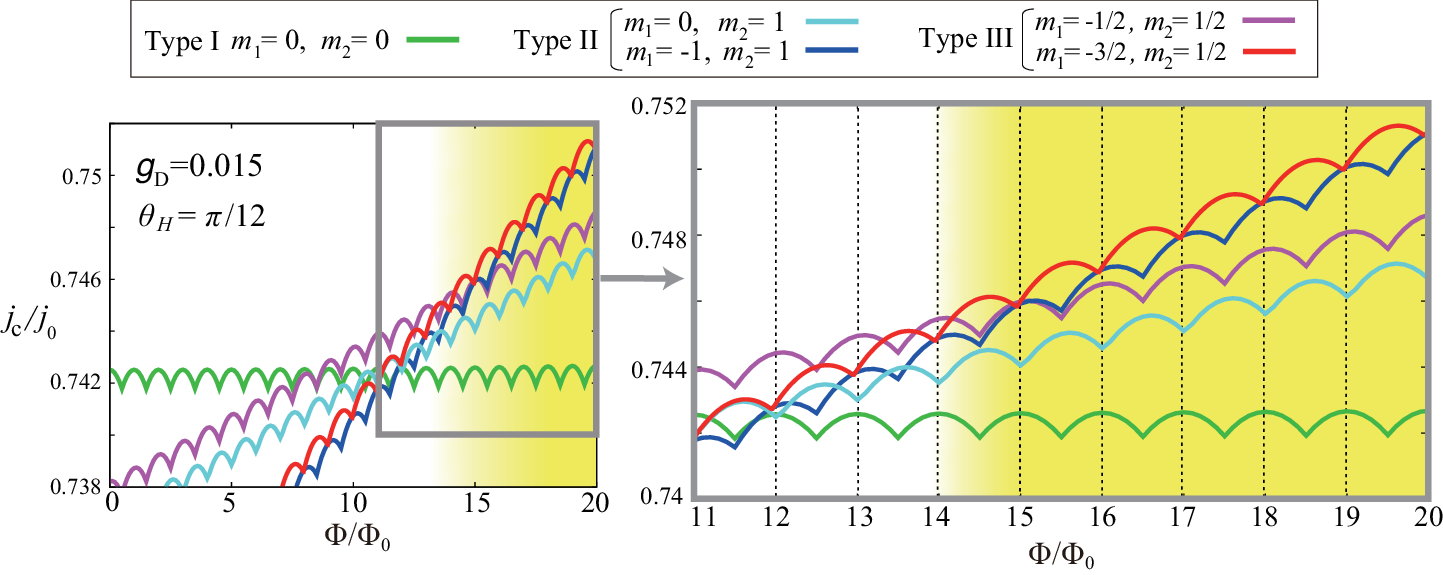}
\caption{The Little-Parks (LP) oscillation in the critical bias current $j_c$ obtained for $g_{\rm D}=0.015$ and $\theta_H=\pi/12$ in the chiral state of $w_{{\bf p},i}=\frac{1}{\sqrt{2}}(\hat{p}_x+i\hat{p}_y)_i$, where other parameters ($T/T_{c0}=0.6$, $R/\xi_0=50$, and $\eta=0.005$) and color notations are the same as those for Fig. 2 in the main text. A right panel shows a zoomed view of the region enclosed by a rectangle in the left panel, where dashed lines indicate peak positions of the LP oscillation for the type-I state realized at low fields (green curve). The HQF state with $m_1=-\frac{3}{2}$ and $m_2=\frac{1}{2}$ is realized at high fields (see the red curve in the yellow region), showing the half-quantum-shifted LP oscillation. \label{fig:supple0}}
\end{figure*}

\section{HQF in the chiral pairing state}
In the case of the chiral pairing state of $w_{{\bf p},i}=\frac{1}{\sqrt{2}}(\hat{p}_x+i\hat{p}_y)_i$, as shown in Fig. 2 in the main text, the type-III HQF state with $m_1=-\frac{3}{2}$ and $m_2=\frac{1}{2}$ becomes more stable with increasing $\theta_H$ for the fixed value of $g_{\rm D}=0.005$, and in the case of $\theta_H=\pi/6$, the HQF state is realized at high fields, showing the half-quantum-shifted Little-Parks (LP) oscillation in the critical bias current $j_c$. For completeness, here, we demonstrate that even for a smaller value of $\theta_H$, the HQF state can be realized for a stronger dipole interaction (larger $g_{\rm D}$). Figure \ref{fig:supple0} shows $j_c$ obtained for $\theta_H=\pi/12$ and $g_{\rm D}=0.015$, where other system parameters are the same as those in Fig. 2 in the main text. Compared with Fig. 2 (b) in the main text, only the $g_{\rm D}$ value is changed with $\theta_H=\pi/12$ remaining unchanged. We note that for these parameter values, type-II and -III states with larger values of $|m_1-m_2|>2$ are basically irrelevant in the field range up to $\Phi/\Phi_0=20$ (for details, see the discussion in Sec. III). One can see from Fig. \ref{fig:supple0} that the HQF state with $m_1=-\frac{3}{2}$ and $m_2=\frac{1}{2}$ (red curve) gives the highest $j_c$ in a high-field region (yellow region), showing the half-quantum-shifted LP oscillation (see the zoomed view). 

\begin{figure*}[t]
\includegraphics[scale=0.48]{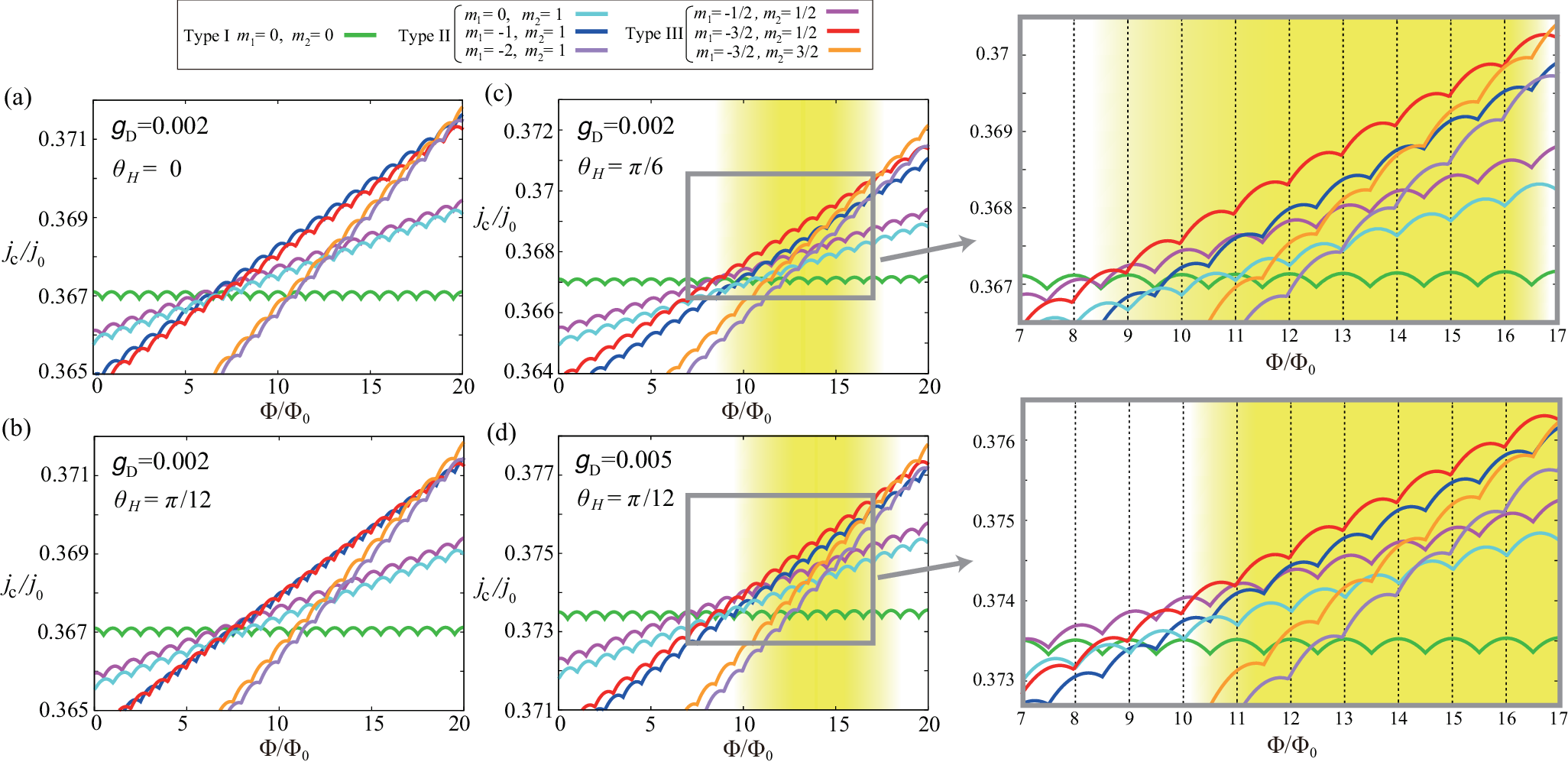}
\caption{The LP oscillation in $j_c$ obtained at $T/T_{c0}=0.6$ for $R/\xi_0=50$ and $\eta=0.005$ in the polar state of $w_{{\bf p},i}=\hat{p}_{z,i}$. (a)-(c) The results for $\theta_H=0$, $\pi/12$, and $\pi/6$, respectively, for the fixed value of $g_{\rm D}=0.002$ and (d) the result for $\theta_H=\pi/12$ and $g_{\rm D}=0.005$, where the color notations are the same as those in Fig. 2 in the main text. In each of (c) and (d), a right panel shows a zoomed view of the region enclosed by a rectangle in the left panel. The HQF state with $m_1=-\frac{3}{2}$ and $m_2=\frac{1}{2}$ is realized at high fields (see the red curve in the yellow region) as in the case of the chiral state. \label{fig:supple}}
\end{figure*}

\section{HQF in the polar pairing state}
As discussed in Sec. I, the difference between the chiral and polar pairing sates consists only in the factors of the coefficients of three terms in the GL free energy, so that qualitatively the same result as that for the chiral sate is naturally expected for the polar state. Figure \ref{fig:supple} shows $j_c$ obtained for the polar pairing state, where the system parameters for (a)-(d) are, respectively, the the same as those in Figs. 2 (a)-(c) in the main text and Fig. 1 in this supplemental material except for the $g_{\rm D}$ value. One can see from Figs. \ref{fig:supple} (c) and (d) that the type-III HQF state with $m_1=-\frac{3}{2}$ and $m_2=\frac{1}{2}$ (red curve) gives the highest $j_c$ in a high-field region (yellow region), as in the case of the chiral state. 

Compared with the chiral-pairing case, the $g_{\rm D}$ value necessary to stabilize the HQF state is smaller in the polar-pairing case. This is simply because for the polar pairing, the absolute value of the coefficient $C_{\rm D}$ is twice larger than that for the chiral state [see Eq. (\ref{eq:coefficient})] and thus, the dipole term $\propto g_{\rm D} |C_{\rm D}|$ stabilizing the HQF state works more efficiently. We note that the $j_c$ value itself in the polar state is smaller than that in the chiral state, reflecting the difference in $C_K$ [see Eq. (\ref{eq:current})]. 

Also, for the present parameter sets, the type-II and -III states with the larger value of $|m_1-m_2|=3$ become relevant on the higher-field side of Fig. \ref{fig:supple}. Such larger-$|m_1-m_2|$ states can appear at higher fields, being irrespective of whether the pairing symmetry is chiral or polar. Since $|\Delta_{\uparrow\uparrow}|^2-|\Delta_{\downarrow\downarrow}|^2$ is roughly proportional to $\eta H$ \cite{LP_KA_prb_22}, the $(|\Delta_{\uparrow\uparrow}|^2-|\Delta_{\downarrow\downarrow}|^2) (m_1-m_2)$ term in ${\bf j}_{\rm tot}$ [see Eq. (\ref{eq:current})] can yield a larger critical current for larger $|m_1-m_2|$ at higher fields (larger $\Phi$). Of course, more complicated ${\bf d}$-vector textures having larger $|m_1-m_2|$ have to pay higher energy cost, so that these textured states are generally unstable at zero field, which is reflected as the downward shift of $j_c$ at $\Phi=0$ from the $j_c$ value for the type-I state of $m_1=m_2=0$.

\end{document}